\documentclass[tightenlines,preprint,superscriptaddress,nofootinbib,showpacs,showkeys]{revtex4} 
\usepackage{graphicx}
\usepackage{dcolumn}
\usepackage{bm}
\newcommand{\be}{\begin{equation}}
\newcommand{\ee}{\end{equation}}
\newcommand{\bee}{\begin{eqnarray}}
\newcommand{\eee}{\end{eqnarray}}
\newcommand{\eq}{\end{quote}}

\begin{document}      
\preprint{PNU-NTG-2/2005}
\title{Determination of the parity of the pentaquark baryons :\\
$\Theta^+$ and $\Xi_5$}
\author{Seung-Il Nam}
\email{sinam@rcnp.osaka-u.ac.jp}
\affiliation{Research Center for Nuclear Physics (RCNP), Osaka
  Unversity, Ibaraki, Osaka
567-0047, Japan}
\affiliation{Deparment of physics and Nuclear physics \& Radiation
technology Institute (NuRI), 
Pusan University, Keum-jung~gu, Busan 609-735, Korea} 
\author{Atsushi Hosaka}
\email{hosaka@rcnp.osaka-u.ac.jp}
\affiliation{Research Center for Nuclear Physics (RCNP), Osaka
  Unversity, Ibaraki, Osaka
567-0047, Japan}
\author{Hyun-Chul Kim}
\email{hchkim@pusan.ac.kr}
\affiliation{Deparment of physics and Nuclear physics \& Radiation
technology Institute (NuRI), 
Pusan University, Keum-jung~gu, Busan 609-735, Korea} 
\date{\today}

\begin{abstract}
We study determination of the parity of the pentaquark baryons in two
different processes. First, we investigate the $\Theta^+$ production via
the $\vec p \vec p \to \Sigma^+ 
\Theta^+$ reaction process which was proposed to   
determine unambiguously the parity of $\Theta^+$. We observe the clear
differences in the total cross sections and the spin 
observable $A_{xx}$. As for the $\Xi_5$ in the $\bar{K}N\to K\Xi_5$
reaction process, the total cross section for  
the positive parity presents about one hundred times lager than that
of the negative one due to the huge destructive interference.   
\end{abstract}
\maketitle
\section{Introduction}
The observation of the evidence of $\Theta^+$ by LEPS
collaboration~\cite{Nakano} motivated by Diakonov 
{\it et al.}~\cite{Diakonov} has intrigued huge amount of the research
activities on hadron physics. Though the quite interesting features of
the pentaquark baryons have been revealed from the experimental
and theoretical approaches so
far, we still have many unknown properties of the baryons which have not been
settled yet. In this report, we study the reaction
dynamics to determine one of the 
unknowns, the parity of the baryons. As
being introduced by Thomas {\it et al.}~\cite{Thomas}, we consider the spin
statistics of the reaction observable in the
$\vec{p}\vec{p}\to\Sigma^+\Theta^+$ process which will shed the light on the
determination of the parity of $\Theta^+$ without model
dependence. We observe clear difference in the size and energy
dependence of the total cross sections depending on the parity of
$\Theta^+$. We
also find the huge destructive interference for the negative 
parity of $\Xi_5$ production from the $\bar{K}N\to
K\Xi_5$ process. The interference provides about one hundred times
smaller total cross 
section for the negative parity $\Xi_5$ than 
that of the positive one. This report is 
organized as follows. In the second section we will present the
results of the  $\vec{p}\vec{p}\to\Sigma^+\Theta^+$ process. The next
section will be given for the $\Xi_5$ production from the $\bar{K}N\to
K\Xi_5$ process. The final section is devoted to summary and conclusions.
\section{The $\vec{p}\vec{p}\to\Sigma^+\Theta^+$ process}
In this section, we consider the $\vec{p}\vec{p}\to\Sigma^+\Theta^+$
process~\cite{Nam4}. This reaction process will provide the
unambiguous method to determine the parity of $\Theta^+$ with the
Pauli principle and the relation of the parity $P=(-1)^{L}$ due to the
following reason. If the initial spin state 
is $S=0$, the final state 
will be dominated by the production of the positive parity
$\Theta^+$ with even $L$. Otherwise, the negative parity 
$\Theta^+$ will be produced when the initial spin state is $S=1$ with
odd $L$. In this way, this method can be addressed as a model 
independent method for the determination of the parity of $\Theta^+$. 

\begin{figure}
\begin{center}
\begin{tabular}{cc}
\resizebox{7cm}{4.5cm}{\includegraphics{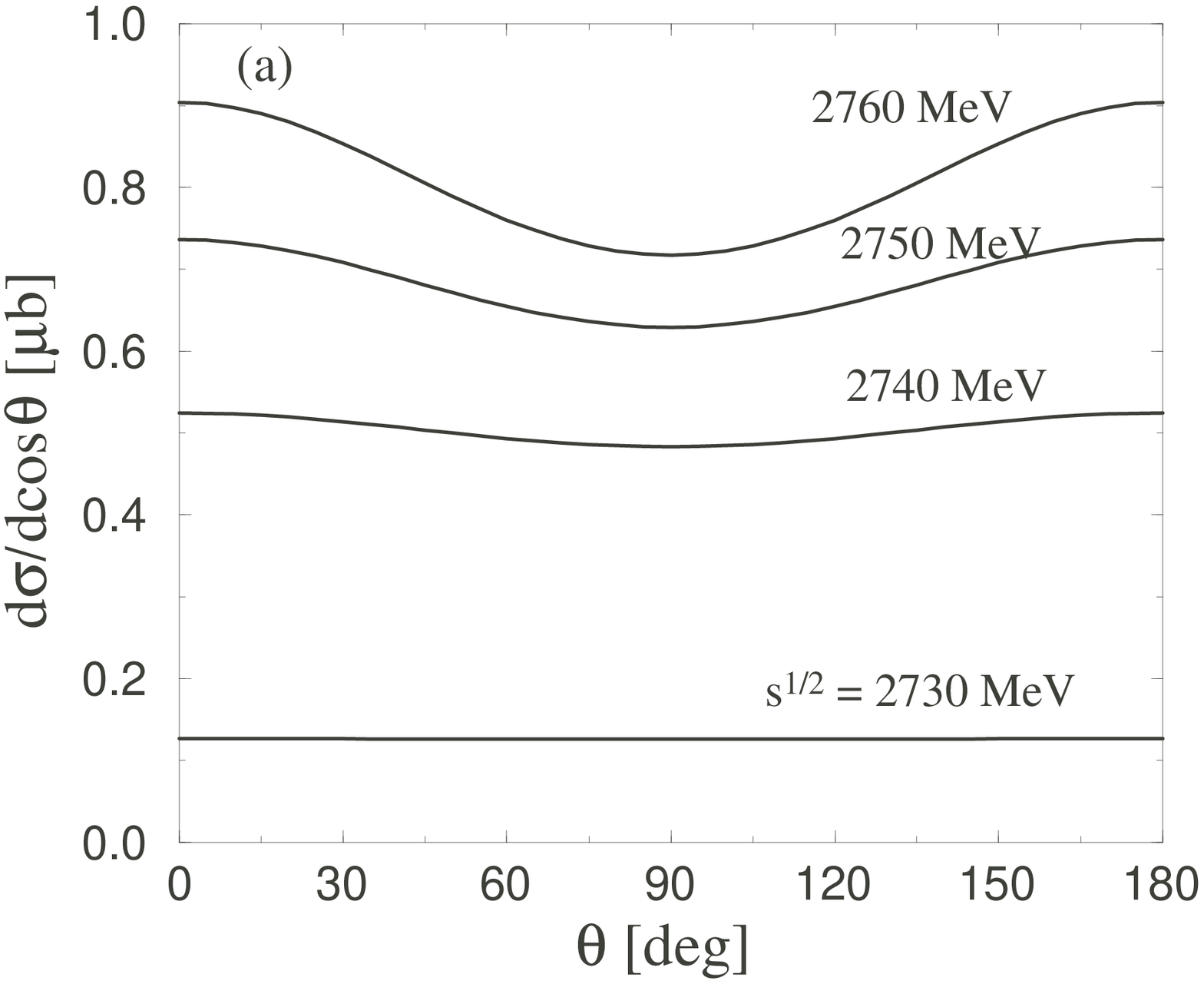}}
\resizebox{7cm}{4.5cm}{\includegraphics{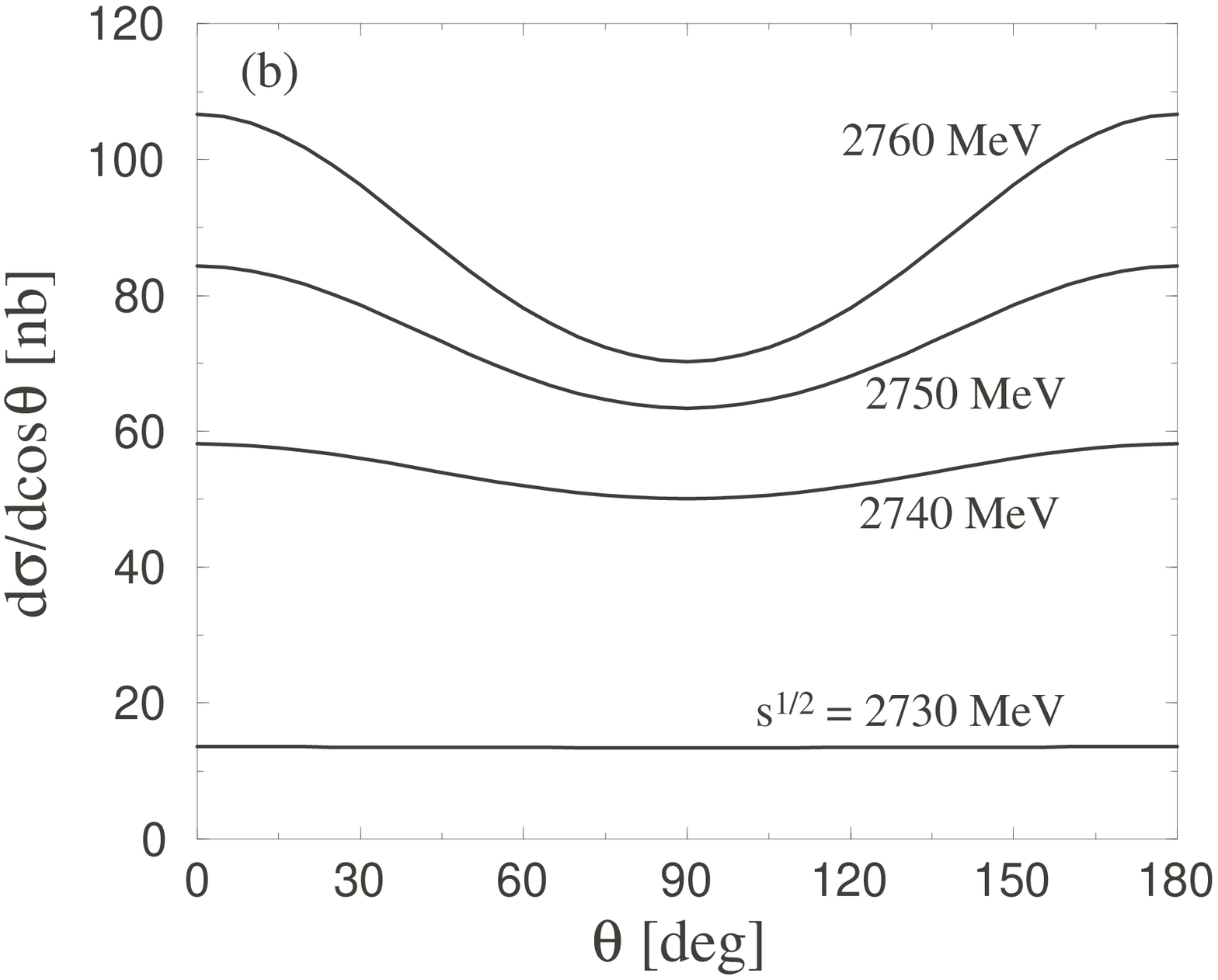}}
\end{tabular}  
\begin{tabular}{cc}
\resizebox{7cm}{4.5cm}{\includegraphics{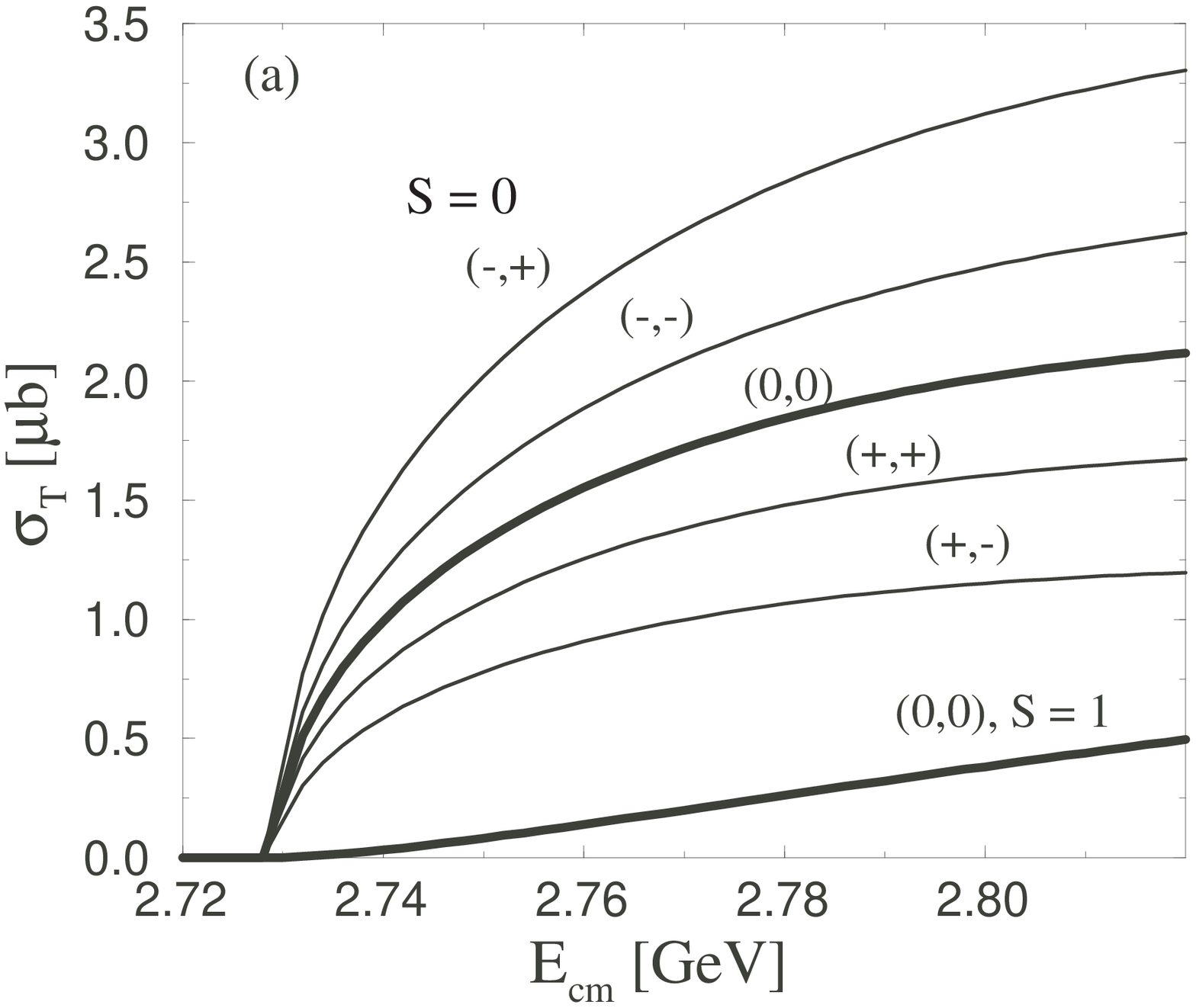}}
\resizebox{7cm}{4.5cm}{\includegraphics{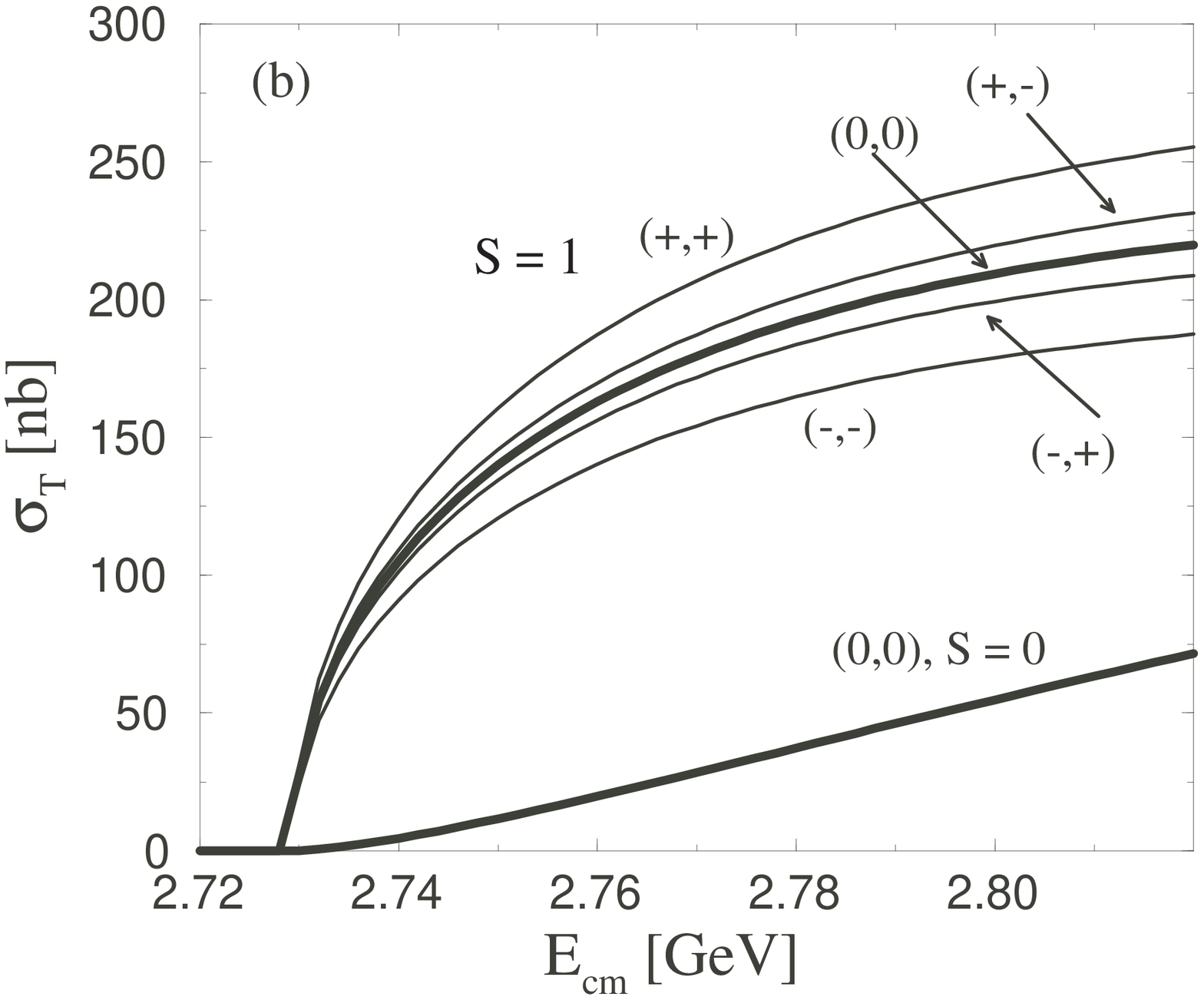}}
\end{tabular} 
\end{center}
\caption{{ Upper panels} : Angular distribution for
the positive parity $\Theta^+$ (left panel) and for the negative
one (right panel). { The lower 
panels} : Total cross sections for the positive parity $\Theta^+$
(left panel)
and the negative one (right panel).}
\label{fig1}
\end{figure}

Now, we perform the Born approximation calculation with the inclusion
of the phenomenological form-factor of dipole type. We assumed one
condition that the final state is dominated by the $S$--wave for the
validity of the method. We have confirmed this by looking at the angular
distribution for the positive and negative parity
$\Theta^+$  as shown in the upper panels of Fig.~\ref{fig1}. The
curves are plotted for 
the several center of mass (CM) energies, $\sqrt{s}=2730\sim 2760$ 
MeV. Near the threshold region, the angular distributions for the two parities are
dominated by $S$--wave contributions. Therefore, one can confirm that
the condition for the model independent method will be 
successfully valid for the low energy region whereas the higher
partial wave
contributions are mixed as the CM energy grows. 

In the lower panels of Fig.~\ref{fig1} we show the total cross
sections for the allowed 
and forbidden channels. We observe the obvious differences in the
order of magnitudes. For the allowed channel, we also show the results
with the different sign combinations of the
$K^*N\Theta^+$ coupling constants which can not be determined by SU(3)
symmetry. As for the forbidden channel, we show only the case without
the $K^*N\Theta^+$
coupling. The parameter dependence seems not so significant. We also note
that the threshold behavior for the allowed channel shows $S$--wave
type ($\sigma\sim (s-s_{\rm th})^{1/2}$) whereas $P$--wave
($\sigma\sim(s-s_{\rm th})^{3/2}$) for 
the forbidden channel. These quite 
different threshold behavior will be critical information when
polarized proton scattering experiments tell the parity of
$\Theta^+$. 

\begin{figure}
\begin{center}
\begin{tabular}{cc}
\resizebox{7cm}{5cm}{\includegraphics{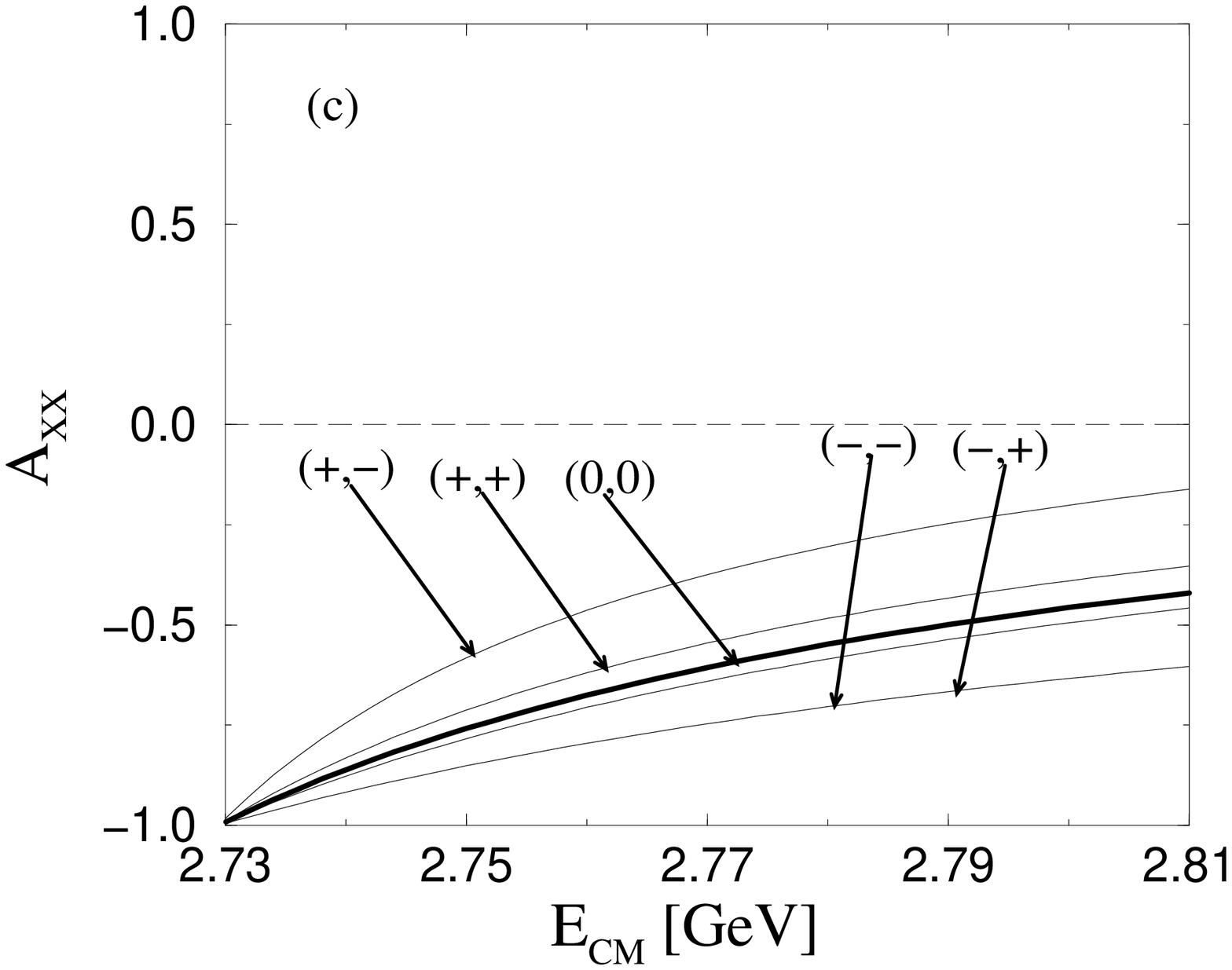}}
\resizebox{7cm}{5cm}{\includegraphics{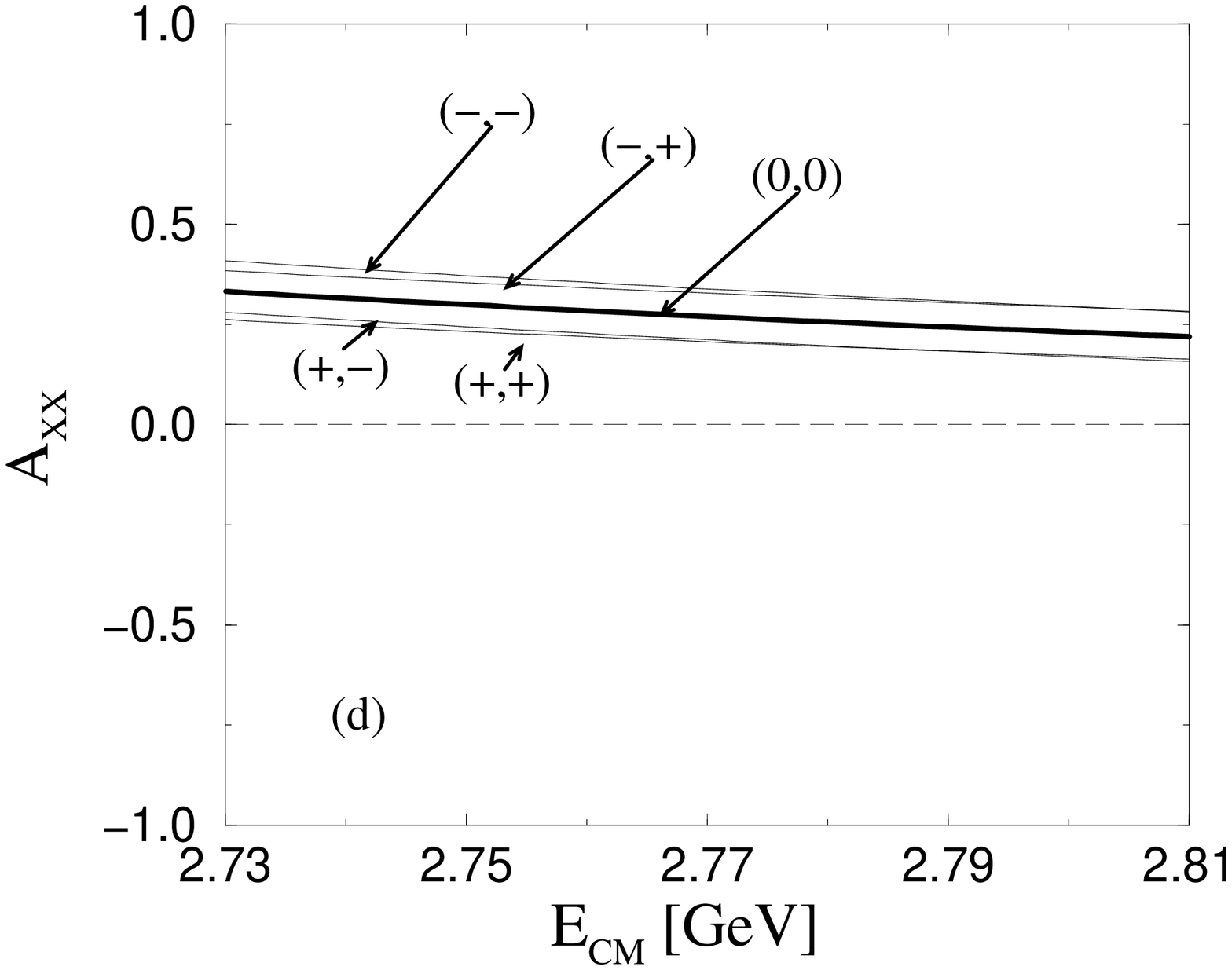}}
\end{tabular}
\begin{tabular}{cc}
\resizebox{7cm}{5cm}{\includegraphics{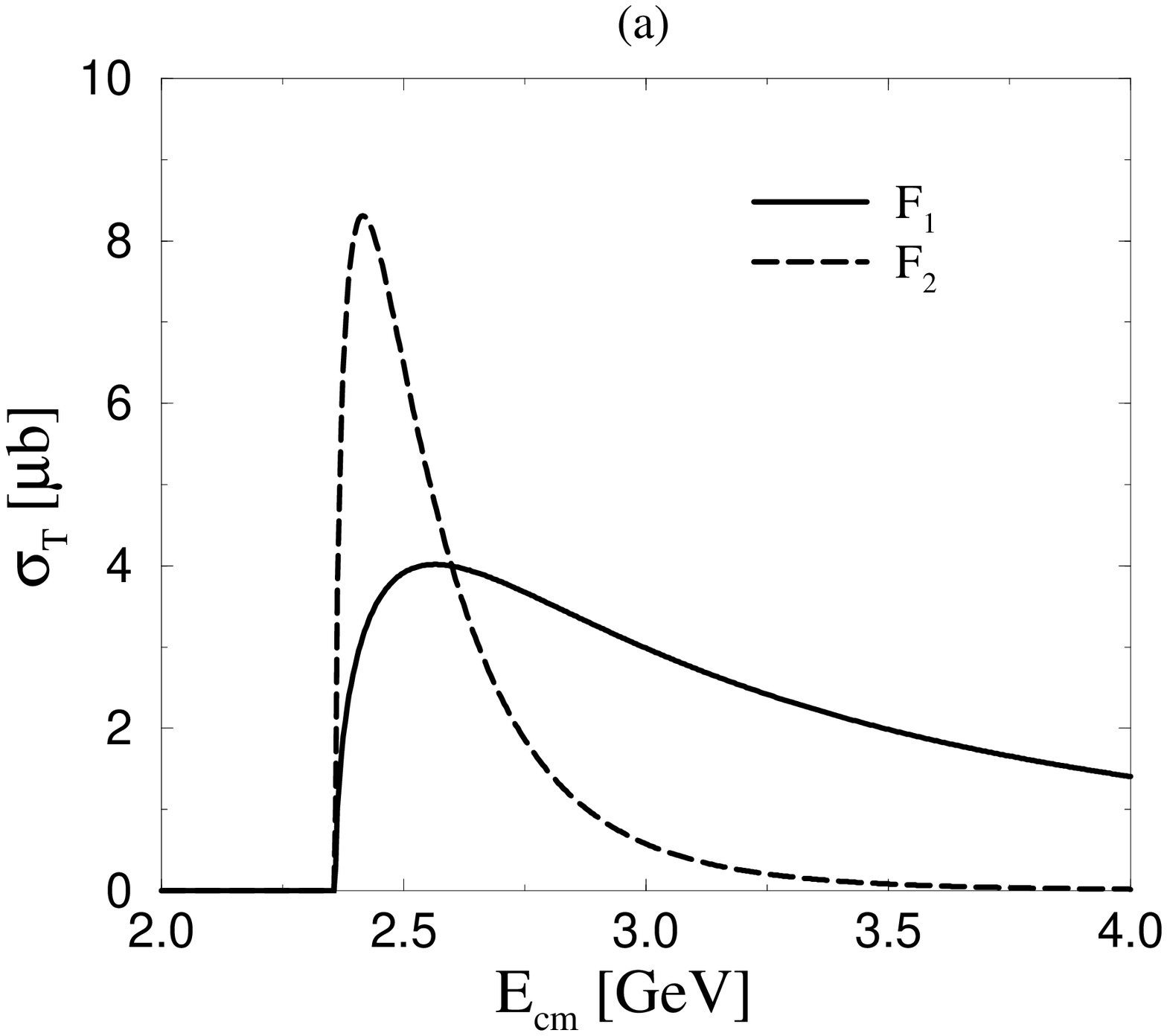}}
\resizebox{7cm}{5cm}{\includegraphics{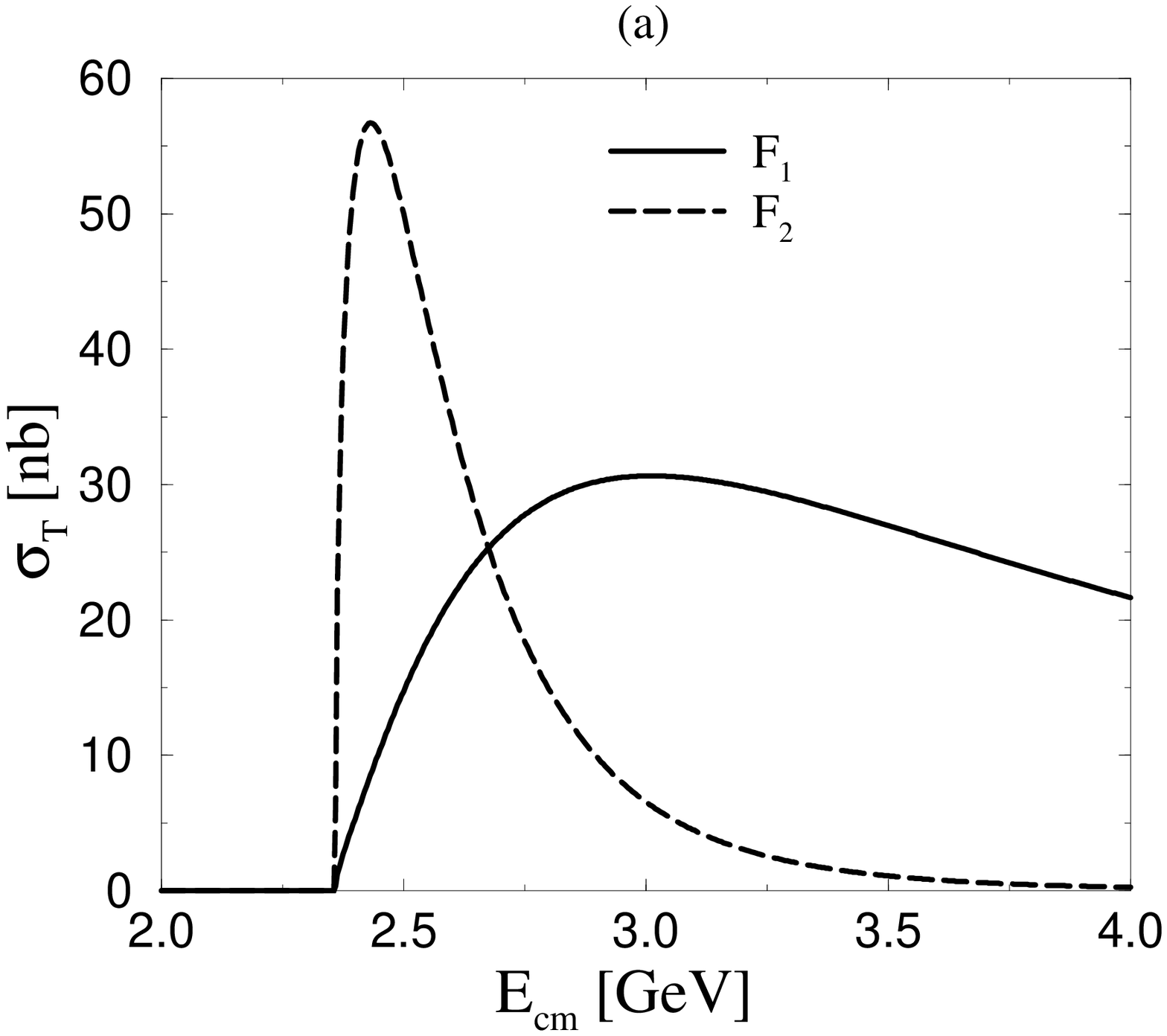}}
\end{tabular}
\end{center}
\caption{{ Upper panels} : $A_{xx}$ for the positive (left
panel) and negative (right panel) parities of $\Theta^+$. { Lower
panels} : Total 
cross sections of the $\bar{K}^{0}p\rightarrow K^{0}\Xi^{+}_5$ process
fpr the positive party $\Xi_5$ (left panel) and the negative one
(right panel).  
}
\label{fig2}
\end{figure}

Hanhardt {\it et al.}~\cite{Hanhart} proposed the spin observable 
$A_{xx}$ which is defined by the following,
\small
\begin{equation}
A_{xx}=\frac{(^{3}\sigma{_0}+^{3}\sigma_{1})}{2\sigma_{0}}-1.
\label{axx}
\end{equation}
\normalsize
As discussed in Ref.~\cite{Hanhart}, this quantity showed clear
separation of the parities. In the upper panels of Fig.~\ref{fig2} we
present the $A_{xx}$ for the two parities of $\Theta^+$. As for the
positive parity, all curves almostly lie in the negative region  and
vice versa for the negative parity $\Theta^+$. This tendency agrees
well with the results of  
Ref.~\cite{Hanhart}. Furthermore, this quantity has an advantage that
the effects caused by the form-factor and unknown parameters will be
reduced since the effects of them appear both in the denominator and
in the numerator at the same time in Eq.~(\ref{axx}), though the cancellation
is not perfect. We note that the spin observable $A_{xx}$ is quite
promising to determine the parity of $\Theta^+$ with less model 
dependence.   
\section{The $\bar{K}N\to K\Xi_5$ process}
NA49 collaboration reported that another pentaquark
baryon $\Xi_5(S=-2,T=3/2)$ with a mass $\sim 1860$
MeV~\cite{NA49}. Though the observation must be
confirmed more carefully, here, we consider the production of the
baryon in  the $\bar{K}N\to K\Xi_5$ reaction process in a simple
Born diagram calculation. It is worth mentioning that there is no $t$--channel
contribution since it is
impossible to change the strangeness by two units by an exchange of a
known meson. The
$K\Sigma\Xi_5$ coupling constant is taken 
from the relation shown in Ref.~\cite{Oh} with $\Gamma_{\Theta^+\to
KN}=15$ MeV. We only consider the $\bar{K}^{0}p\to K^{0}\Xi^{+}_5$
isospin channel since the difference due to other isospin channel is
small. In the lower panels of Fig.~\ref{fig2} we show the total cross
sections for the process 
with the two parities of $\Xi_5$. In order to check the model
dependence, we consider two different types of the form-factors, $F_1$
and $F_2$~\cite{Nam6}. Though we observe quite sizable model
dependence from the form-factors, we would like to emphasize that huge
difference in the order of the magnitudes for the two parities which
is about factor one hundred. This result is an interesting
feature of the $\Xi_5$ production which will be helpful to give a
qualitative determination 
for the parity if experimental data will be available.      
\section{Summary and conclusions}
We performed calculations for the two pentaquark
production processes, $\vec{p}\vec{p}\to\Sigma^+{\Theta^+}$ and
$\bar{K}N\to K{\Xi_5}$ by using the Born approximation. The
main purpose 
for these calculations was to provide reliable methods and
information to determine the parities of the two pentaquark
baryons. We also considered model dependence of the cross sections
through unknown coupling constants and form-factors. We found quite
useful information to determine the parity of $\Theta^+$ in the
polarized proton scattering. Since this reaction is less model
dependent, the results of this work is 
quite promising for the determination in future experiments. Also, the
interesting feature in the $\Xi_5$ production reaction process can be
a good milestone at the first stage of the analysis in experiments. As
planed by the COSY-TOF collaboration~\cite{COSY-TOF}, the polarized
proton scattering 
will be done in near future with our expectation for the determination
of the parity of $\Theta^+$.        

\end{document}